# Beam Measurements and Machine Learning at the CERN Large Hadron Collider

P. Arpaia[1], G. Azzopardi[2], F. Blanc[3], X. Buffat[2], L. Coyle[2,3], E. Fol[2,4], F. Giordano[1,2], M. Giovannozzi[2,*], T. Pieloni[3], R. Prevete[1], S. Redaelli[2], B. Salvachua[2], B. Salvant[2], M. Schenk[2], M. Solfaroli Camillocci[2], R. Tomás[2], G. Valentino[5], F.F. Van der Veken[2,5], J. Wenninger[2]

[1] Dipartimento di Ingegneria Elettrica e Tecnologie dell'Informazione (DIETI), Università degli studi di Napoli Federico II, 80125 Napoli, Italy

[2] Beams Department, CERN, Esplanade des Particules 1, 1211 Geneva 23, Switzerland

[3] Ecole Polytechnique Federale Lausanne, 1015, Lausanne, Switzerland

[4] Johann Wolfgang Goethe Universität, Max-von-Laue-Strasse 9, 60438 Frankfurt, Germany

[5] University of Malta MSD2080, Msida, Malta

[*] Corresponding author: massimo.giovannozzi@cern.ch

## Introduction

Particle accelerators are among the most complex instruments conceived by physicists for the exploration of the fundamental laws of nature. Of relevance for particle physics are the high-energy colliders, such as the CERN Large Hadron Collider (LHC), which hosts particle physics experiments that are probing the Standard Model predictions and looking for signs of physics beyond the standard model.

The particle beams circulating in the colliders need to be controlled in an accurate way with beam instrumentation key to the performance of a collider. The amount of data generated by beam instrumentation is so large that Machine Learning (ML) is making its way into the domain of Accelerator Physics, with several laboratories worldwide devoting intense efforts in this domain.

The power of these tools has been exploited for advanced analysis of colliders data since decades, but it is only in recent years that ML techniques have found applications in the field of Accelerator Physics. The first attempts to use these techniques date to a few decades ago and dealt with beam diagnostics and beam control systems [1,2]. However, some sizeable progress has been made only recently (see, e.g. [3-8] and references therein). Nowadays, there is a general agreement within the Accelerator Physics community of the need and usefulness of ML techniques, which resulted in the publication of a white paper reviewing the state-of-the-art and presenting recommendations to encourage the development of such techniques in Accelerator Physics laboratories [9].

CERN has recently started focused efforts oriented towards ML techniques for beam dynamics studies at the Large Hadron Collider (LHC). The LHC complexity, in terms of number of operational systems, size of data collected, variety of beam dynamics configurations and beam behaviours, is such that ML becomes an efficient tool for data analysis. Indeed, a wide spectrum of applications from beam measurements and machine performance optimisation to analysis of numerical data from tracking simulations of nonlinear beam dynamics will be reviewed in this paper, paying attention to future developments in view of projects like the Future Circular Collider (FCC) [10] whose complexity will far exceed that of the LHC.

The process of building a mathematical model built upon sample data (*training data*), makes the core of ML, which aims for predictions or decisions be made without explicit programming required [11], which makes ML so appealing in particular when there is a large amount of data. ML includes a few learning paradigms, such as Supervised Learning (SL), Unsupervised Learning (UL), and Reinforcement Learning, and deals with tasks such as classification, regression, clustering, anomaly detection, dimensionality reduction, and reward maximisation [12].

To train a mathematical model to successfully achieve a particular task requires a few complex steps for ML, which include data collection and curation, feature (input) engineering, feature selection and dimensionality reduction, model hyper-parameter optimisation, model training, with performance evaluation being at the end of this process.

In the SL approach, training of ML algorithms is carried out on labelled data sets where a ground-truth output exists for each input. In UL [13], however, no ground-truth output is available, and the algorithms aim at discovering structure in the dataset. Reinforcement Learning is typically used in control applications where the goal is to execute a series of actions to maximize a given reward, such as achieving a given beam parameter.

**Overview of the LHC ring and optics measurement**

Figure 1 shows the schematic layout of the LHC ring (see [14] for additional detail).

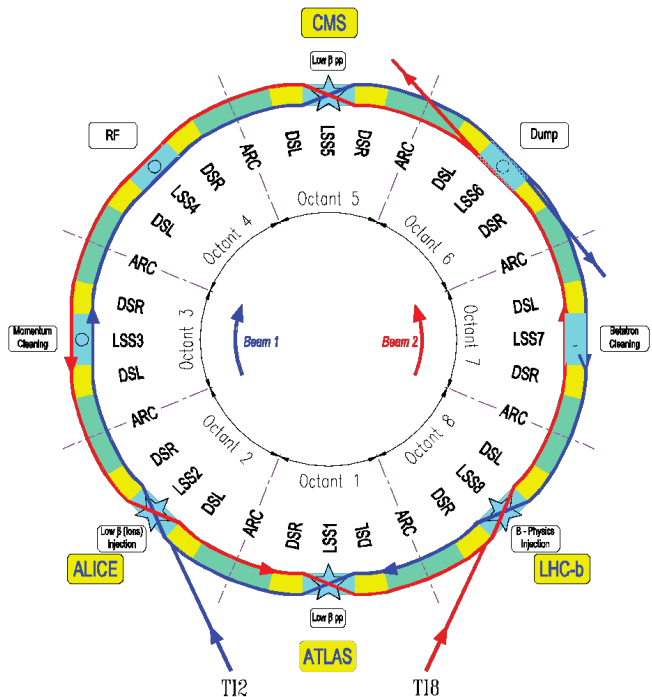

*Figure 1: Layout of the LHC ring (from [21])*

The eight-fold symmetry is visible together with the main function of each long straight section. Note the RF accelerating systems share the straight section with some key beam diagnostic devices, like transverse and longitudinal beam profile monitors and beam current transformers. During the LHC Run 2 (2015-2018), proton and ion (of atomic number Z) beams were accelerated from 450 Z GeV to a maximum of 6.5 Z TeV (see, e.g. [15]).

The study and optimisation of the linear optics [16] has been a priority for the tight link with the collider performance. However, even an excellent theoretical optics is of little use if excellent measurements and corrections techniques are not at hand. For this reason, strong efforts have been devoted in recent years to the development and improvement of the techniques for measurement and correction of the linear optics, with outstanding results [17-20] that have been of paramount importance for achieving the LHC excellent performance. Further improvements would require tackling two major points: devising approaches to detect faulty Beam Position Monitor (BPM) measurements to exclude them from the further analyses and building effective models to describe how the magnetic field errors distributed along the accelerator circumference influence the optics measurements. ML techniques provide tools to deal with both points and have been the subject of active research efforts in recent years.

Both SL and UL paradigms are applied to measurements of linear optics and its correction at the LHC. Supervised methods are used to build regression models with the goal of reconstructing errors of individual magnets from optics disturbances generated by the errors, whereas correction techniques available to date compute circuits strength settings to compensate the measured optics deviations from the design values. The initial results discussed in [21-24] show very clearly the capability of ML-based regression models [25-29] to predict individual quadrupolar errors. More than this, they can boost the quality of optics corrections, as they provide insight into the sources of local errors in the accelerator. Some novel findings are presented in [30] and cover cases where the data are from numerical simulations and beam measurements.

The detection of faulty BPMs in optics measurements, responsible for nonphysical outliers in the value of the optical functions derived from BPM data, is better performed by means of UL tools. A decision-tree-based Isolation Forest algorithm (IF) [31] has considerably improved the measured data used to evaluate optical functions, with an important side effect of minimising the human effort to perform the cleaning of measured data.

Anomaly-detection techniques have also been applied to cleaning measured data with a comparison to clustering techniques, as presented in [32], while extensive studies on cleaning techniques for the LHC optics measurements, comprising recent progress, and plans for the future, are discussed in [33].

## LHC collimation system commissioning with beam

The LHC equipment must be protected from any damage or down-time caused by beam losses, for which a robust collimation system is used. It comprises around 100 collimators located along the 27 km ring, each made of two parallel absorbing jaws. The four jaw ends are individually movable by means of dedicated stepper motors, which totals about 400 degrees of freedom.

Collimator settings are determined following a beam-based alignment (BBA) procedure established in [34], which is used to estimate the beam centre and size at the collimator locations. This procedure moves collimator jaws separately towards the beam halo, whilst monitoring the measured beam loss signal on a dedicated Beam Loss Monitoring (BLM) device positioned immediately downstream. A collimator is aligned when both jaws are centred around the beam orbit, and this occurs once a particular pattern of spike has been detected in the BLM signal for each jaw. The BBA is semi-automated [35], but the manual detection and classification of the alignment spikes required is a tedious and time-consuming procedure that has to be carried out by system experts.

During the initial beam commissioning period taking place each year, the collimators have to be aligned so to ensure their correct setup: at injection energy 79 collimators are aligned, while 75 collimators are aligned at top energy.

The campaigns of collimator alignment involve continuously moving the jaws towards the beam, while ignoring non-alignment spikes, until the observation of a clear alignment spike. The alignment spike shown in Fig. 2 indicates that the beam halo has been touched by the moving jaw. It is made of a steady-state signal, followed by a loss spike, with temporal decay of losses, and then a steady-state signal following the spike. The second steady state, featuring losses larger than the first one, is due to the continuous scraping of halo particles when the jaw are in fixed positions, which is induced by a transport mechanism towards high amplitude like a diffusive process. Any other spikes not following this pattern are classified as non-alignment spikes. Achieving a reliable alignment implies the capability of correctly identifying the spikes in the BLM signal, and in a typical alignment campaign, hundreds of such spikes require analyses.

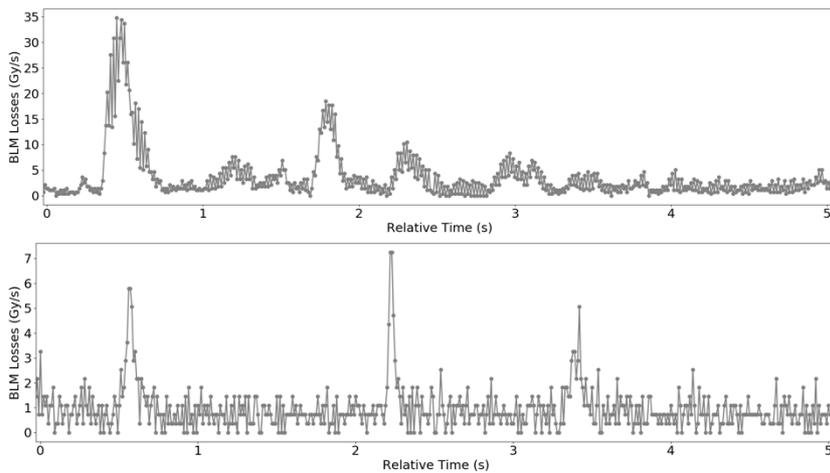

Figure 2: Representative 100 Hz BLM signals with a clearly visible alignment spike, the first one in the series of spikes, (top) and spurious spikes (bottom), following inward collimator movements at about t = 0.5 s (from [43]).

The spike recognition process can be seen as a classification problem, and training of ML models was performed to discriminate the two spike patterns in the BLM losses [37]. Data were collected during 11 semi-automatic collimator alignment campaigns carried out in 2016 and 2018 (injection and top energy). In total, 6446 samples were catalogued, out of which 4379 represented alignment spikes whereas 2067 spurious spikes. The information acquired in alignment campaigns consisted of 100 Hz BLM data and collimator jaw positions stored at a frequency of 1 Hz.

Fourteen manually engineered features describing the shape of the spikes observed in the BLM signal (see [37] for a detailed description of the features considered) were extracted from this data set and the most relevant were selected by means of the strength of association between each pair of variables, determined using the Spearman correlation. Five different ML models analysed the feature-selection stage, to check how they ranked the importance of each feature. These models were individually trained exploiting all features, and the features were ranked in ascending order, depending on their relevance. A Sequential Forward Selection algorithm (SFS) [38] was applied to distil the most appropriate features using the best hyper-parameters. The SFS algorithm checks all combinations of features by introducing one at a time and retaining the best one for future combinations.

The five topmost characteristics [36] were applied for training and comparing six ML models for binary classification, namely Logistic Regression, Neural Network, Support Vector Machine, Decision Tree, Random Forest, and Gradient Boost. False detection of an alignment spike is to be considered worse than not detecting an alignment one, hence the main performance metric used was precision. Figure 3 shows the distribution of precision obtained by each model and their Ensemble, i.e. a new model in which the others are gathered together by a mechanism of majority vote. The differences in the mean precision of the results obtained from each model are included in Fig. 3, a different colour implying that the mean significantly differs.

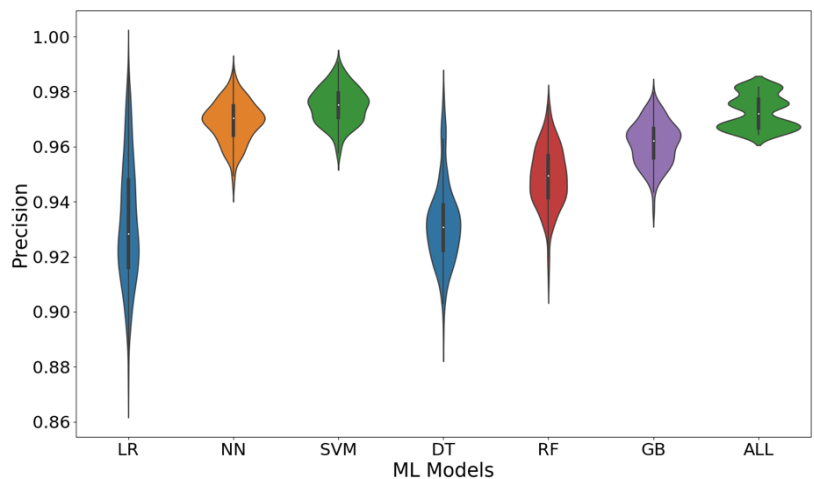

*Figure 3: Precision distribution for each model considered and their Ensemble (LR: Logistic Regression, NN: Neural Network, SVM: Support Vector Machine, DT: Decision Tree, RF: Random Forest, GB: Gradient Boost, ALL: Ensemble) (from [36]).*

The Support Vector Machine achieved the best precision, featuring a mean like the Ensemble. The Ensemble model was added to the BBA software [39], including the necessary threshold-selection algorithm [40] and crosstalk analysis [41], which transformed the semi-automatic alignment into a fully automatic one. This new alignment software was successfully used throughout 2018 LHC operation. The first version was tested during beam commissioning, with the collimators of the two beams being automatically aligned in sequence, at injection and top energy. The possibility to align the collimators of the two beams in parallel was tested in a machine development (MD) study, with another MD scheduled to probe the parallel fully automatic software including angular alignments. The alignment time of the collimators at injection was reduced by 71.4%, with respect to the semi-automatic alignment in 2017 (from nearly 3 hours to just 50 minutes) [42], as shown in Fig. 4. This represents a major step in terms of operational efficiency enhancement and will become the default software for starting the LHC in 2022.

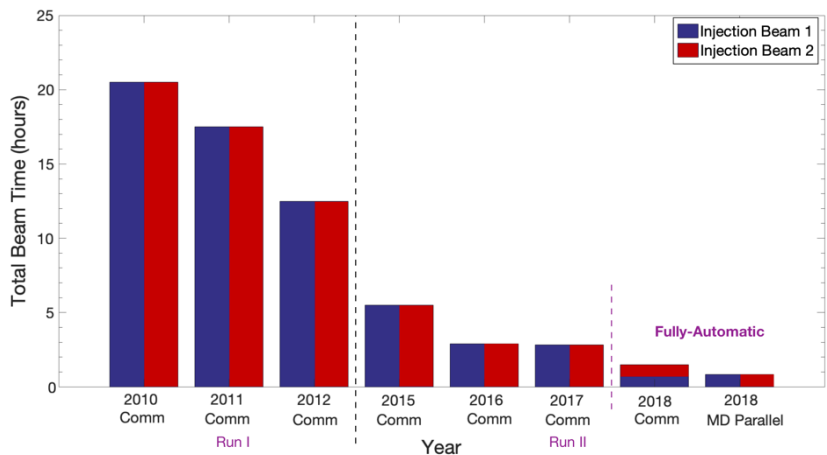

*Figure 4: Evolution of the time required for the collimators alignment at injection (from [42]). The first 2018 data indicate that the alignment had been carried out in sequence for the two beams, whereas the second 2018 data show parallel alignment of the two beams.*

**Beam lifetime and losses optimisation**

Using ML to build a rigorous model of particle losses occurring in the machine would offer valuable insights into the detail of the beam dynamics, which would help enhance the LHC performance, also allowing the exploration of ML applications for future colliders, such as the CERN FCC [10].

The main aim consists of developing an approach capable of identifying the set of operational parameters that provide the maximum beam intensity lifetimes under the constraint of a given, specific machine configuration [43]. The ultimate goal is the comparison of the model obtained from experimental data, which would be a data-driven surrogate model, against results from particle tracking simulations: an efficient optimisation algorithm could then be applied to establish the best parameters for operation.

This problem has been tackled with a SL approach: the model output is the beam lifetime, while the physical accelerator parameters, such as tune, chromaticity, and magnet currents are the inputs. The used data cover an entire operational year, but only a reduced subset of the complete machine cycle, corresponding to the end of the injection plateau just prior to launching the energy ramp, is considered: such a severe data reduction will be progressively removed in future analyses.

Several SL models were trained and compared, and the best performance was achieved with a Gradient Boosted Decision Tree model [44]. The extraction of the optimal machine configuration from the trained lifetime response was made by using an off-the-shelf simplex optimiser [45]. An important observation is that the input data distribution is not optimal, as LHC operation reproduces strictly a given parameter set in every cycle to prevent performance loss due to unconventional machine settings. Therefore, the surrogate model trained on available data exactly represents the operational machine configurations but features a scarce predictive power for unconventional machine setups, as these are underrepresented, if not absent, in the data set used to build the surrogate model. Dedicated experimental sessions were therefore performed to collect data representing unconventional settings of the LHC. Machine configurations were varied and correspondingly, tune scans were carried out [46], and the collected data are used to test and complement the current model. It is worth noting that beam instabilities increased the beam emittances, thus affecting the machine performance. These instabilities are not considered by the current model, another weakness of the existing setup. If this issue is ignored and the goal restricted to the problem of optimising the lifetime, agreement with the optimal lifetime region of the measured tune diagram is found, as seen in Fig. 5, where the maximum measured lifetime (red circle) is shown together with the optimal value from the surrogate model (blue circle). In particular, the model is capable of detecting the optimal regions, reaching an improvement of the lifetime of a factor of two, although it misses the maximum. One possible reason for this is the rather large distance of the nominal working point from the optimised working point showing the maximum lifetime, with the model seeing relatively few of such configurations in the training data.

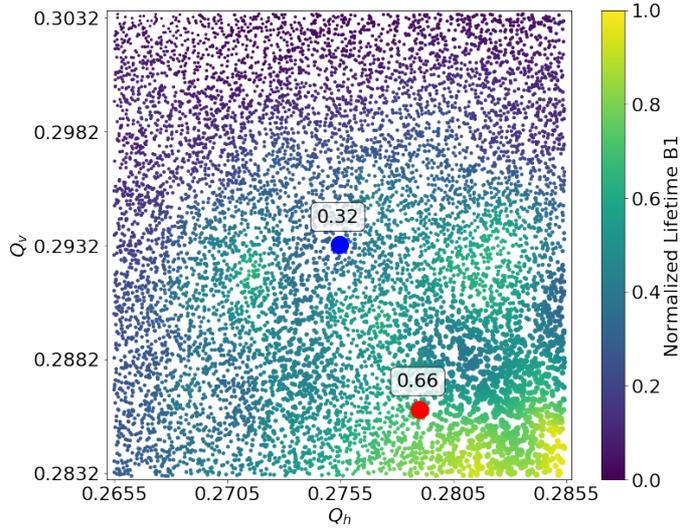

*Figure 5: Measured normalised Beam 1 lifetime as a function of the LHC transverse tunes. Blue dot: Nominal working point. Red dot: Lifetime-optimised working point as computed by the surrogate model. The prediction (blue circle) is close, but not equal, to the measured maximum lifetime (red circle), the absolute minimum and maximum values of the beam lifetime being 1.05 h and 32.7 h, respectively (from [36]).*

**Detection of beam instabilities from stored data**

Beam instabilities due to collective motion can generate to a severe degradation of beam quality and to a reduction in collider luminosity. It is therefore crucial to gain insight into the conditions that lead to instabilities, in view of finding appropriate mitigation measures. A few dedicated measurement devices are installed at the LHC for this purpose, and here we focus on the data collected through the observation box (ObsBox) of the transverse damping system [47].

This device keeping a rolling buffer of data coming from several BPMs. When triggered, either automatically or manually, the ObsBox writes the entire buffer to disk. Here, the focus is on data saved upon automatic detection of an instability. The data saved in this manner is of very high resolution, i.e. it comprises bunch-by-bunch, turn-by-turn, and transverse beam position information along the machine cycle for all beam modes and fill types, with around 4 TB of data accumulated so far. The data set presented here was collected between 5[th] of September 2017 and 3[rd] of December 2018 and includes a total of 36196 triggers. However, in most of the data was found no signs of beam instabilities. A procedure to identify unusual beam oscillation patterns in the data and cluster the various types of signals together was therefore developed to speed up the data analysis.

Before any clustering can be performed, the false triggers need to be removed from the data. This step is equivalent to an anomaly-detection problem with the false triggers being the nominal samples and the instabilities the anomalies. Data filtering is conducted by the extraction of several features present in the data [36], on which Principal Component Analysis (PCA) [48] is carried out. Such a PCA space can have a dimension as low as 4, while still describing 93% of the variance of the extracted attributes, as shown in Fig. 6.

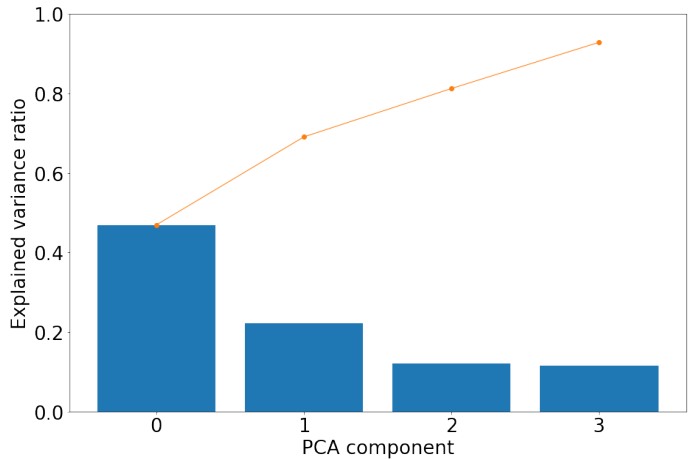

*Figure 6: Principal Component Analysis of ObsBox data collected with automatic triggers. The histogram shows the explained variance of each PCA component, while the curve represents the cumulative explained variance (from [36]).*

A standard Isolation Forest (IF) [31] algorithm can be applied to the PCA space to identify the anomalous samples.

In this way, the trained IF can perform the filtering of the false triggers from the ObsBox data. Then, more computationally intensive algorithms can be run on the predicted anomalous samples from the entire set of ObsBox data in view of more detailed analyses. At this stage, each bunch is considered as independent, implying that phenomena like coupled-bunch instabilities are disregarded, which greatly simplifies the cluster identification as this then applies to a univariate time series. After having determined the clustering one can proceed with a Hierarchical Clustering Algorithm [49] to identify iteratively and link clusters in a form of a dendrogram.

The experience gathered so far is that signals with similar patterns are, for the most part, clustered together, although some issues have been observed, linked with the fact that the designed clustering algorithm does not allow for a partial match of the time series. In summary, there is an enormous potential for ML techniques to correctly identify false triggers, thus allowing for a more efficient and detailed study of the relatively few data sets that contain instability data.

**Pressure readings and heating detection**

Charged particles stored in a high-energy, high-intensity accelerator ring may generate heating in the surrounding equipment due to electron cloud effects [50], particles lost on the beam surroundings [51], synchrotron radiation effects [52], and beam-induced RF heating due to impedance effects [53]. In a high-vacuum environment, temperature increase may lead to outgassing [54] that appears as a pressure increase in vacuum gauges, Beam-induced heating can be directly measured by means of temperature probes. However, standard vacuum monitoring is more widespread and systematic than temperature monitoring, e.g. more than 1200 vacuum gauges are distributed along LHC ring circumference [56]. Pattern analysis of the vacuum gauges readings after each LHC fill to identify abnormal behaviour is a time-consuming and heavy task. Furthermore, there is no reliable technique to convert the vacuum measurements into equivalent temperature values.

The pressure readings produced by the vacuum gauges are analysed by a newly developed automatic classification algorithm in view of detecting heating patterns from an anomalous pressure increase. Figure 7 shows the time evolution of the pressure of a Beam 2 vacuum gauge situated in Sector 4-5 of the LHC, in the vicinity of the stand-alone magnets D4 and Q5. Sudden changes in pressure reading are clearly visible and may be linked with outgassing generated by a temperature rise. Note we disregard the underlying physical phenomenon inducing the temperature rise as only its effects are considered.

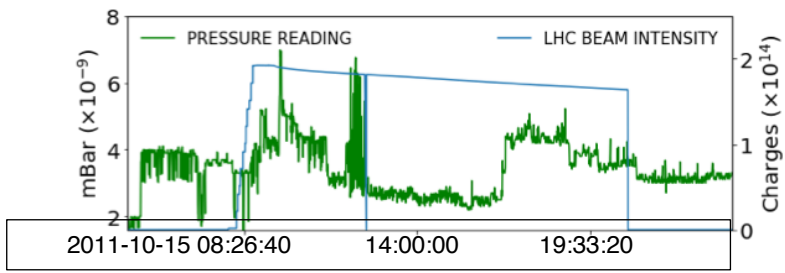

*Figure 7: Typical data (pressure readings and beam intensity) used to estimate beam heating effects from pressure readings of a Beam 2 vacuum gauge situated in Sector 4-5 of the LHC, in the vicinity of the stand-alone magnets D4 and Q5, during a physics fill of the 2015 run. The sudden pressure increase may indicate a beam-induced heating effect close to the vacuum gauge (from [36]).*

To speed up finding the anomalous gauge readings, the classifier selects a subset of all gauges containing such signals, i.e. the classifier should aim at a high *recall* score, defined as the fraction of true positives detected from the total number of positive cases.

More than 700 vacuum gauge readings were labelled as anomalous through human expert supervision, and a data set containing 700 time series of 3000 time steps each was created. A PCA was then performed resulting in the retention of only 12 features that retain no particular physical meaning, but explain 99.9% of the full data set variance, hence without leading to a significant loss of information. A K-Nearest Neighbour Classifier (KNN) followed by a Multi-Layer Perceptron (MLP) were then trained on the resulting data set containing only these 12 features (see [36] for details).

The performance of the KNN and MLP classifiers was evaluated by means of a 4-fold cross-validation approach [57], applied when training each model. The 4-fold technique implemented Stratified splitting [58], the folds being built by preserving the percentage of samples for each class.

Figure 8 (left) shows the results of the parameter set scan maximising the *recall* for the KNN. The red dots, corresponding to *recall*=1 in the KNN classifier, are parameter sets for which the algorithm overfits the training data, i.e. it behaves extremely well on training data, but poorly on the test data.

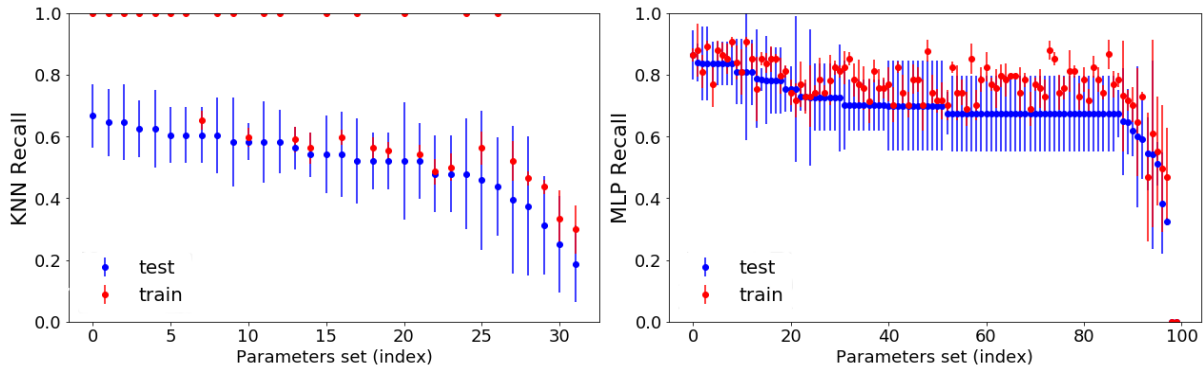

*Figure 8: Recall scores of the KNN classifier (left) and of the MLP (right). The red dots at recall=1 in the left plot are parameters set for which the algorithm overfits the training data (from [36]).*

At index 7 in the plot, one can find the best parameter set, i.e. the first set of parameters that does not overfit the training set, for which *recall*=0.60 ± 0.09. The results of the MLP scan over the network parameters is shown in Fig. 8 (right). The best parameter set is achieved at index 0, corresponding to a network made of 2 hidden layers with 176 neurons per layer. For the neural network *recall*=0.86 ± 0.10, an improvement of 35% with respect to the KNN case.

Hence, a rather simple Neural Network features interesting *recall* scores, thus motivating the test of more refined ML techniques on this task, such as Convolutional Neural Network and ensemble methods (see [36]).

**Digression: ML learning applied to simulated measurements of dynamic aperture**

This section is devoted to the presentation and discussion of applications of ML to numerical simulations, hence not beam measurements, of nonlinear beam dynamics. Among various concepts that can be used to describe nonlinear beam dynamics, that of Dynamic Aperture (DA) is particularly useful. It is the radius of the smallest sphere inscribed in the connected volume in phase space where bounded motion over a given time interval occurs [59]. Tracking simulations are used to estimate the DA by probing the evolution of a certain set of initial conditions, usually uniformly distributed in polar co-ordinates in normalised physical space. This procedure is repeated for different realisations of the magnetic field errors (the so-called seeds) for a certain accelerator model, and an average DA can be defined

$$DA_{\text{ave}} = \frac{1}{N_{seed}} \sum_{i=1}^{N} \int d\theta \, r_i(\theta),$$

where $r_i(\vartheta)$ stands for the last stable amplitude for seed $i$ in the direction of angle $\vartheta$. For design studies, a conservative estimate is preferred, and the DA is estimated as

$$DA_{min} = r_{i,j}(\theta), \quad \text{where } 1 \leq i \leq N_{seed}, \; 1 \leq j \leq N_{angle}$$

and $r_{i,j}$ represents the last stable amplitude for the $i$th seed and $j$th angle. $DA_{min}$ might be strongly influenced by outliers, and efficient automatic tools to tackle the identification of outliers would be very useful. Figure 9 shows examples of DA plots, generated for two LHC configurations, using sixty seeds, eleven angles, and $10^5$ simulated turns. The left plot refers to the optics version 1.3 for the High-Luminosity LHC at top energy, with $\beta^*$=15 cm, Q'=15, and strong powering of the Landau octupoles, with no beam-beam effects. The right plot refers to the optics configuration of the LHC for the 2016 proton run at injection energy, Q'=8, and strong Landau octupoles to combat electron-cloud effects.

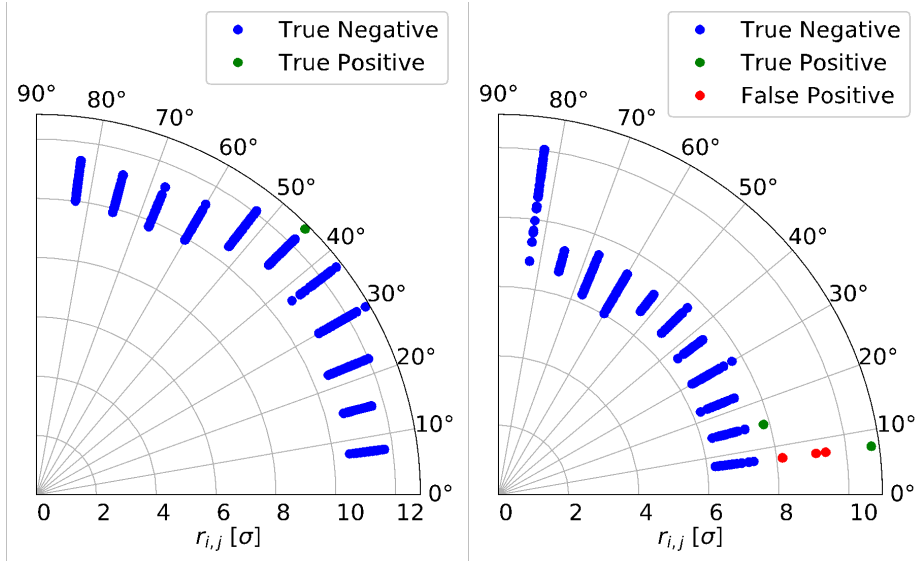

*Figure 9: DA simulations for the v1.3 HL-LHC configuration (left) and 2016 LHC configuration (right). Initial 4D co-ordinates like (x, 0, y, 0) are used, and a polar scan is carried out on (x, y), only. The markers represent the results of the sixty seeds. Left: Example of DA computations where an outlier is correctly flagged (green). Right: Examples of false positives (red). False positive cases are of less concern in this case, as they have an impact on the determination of the maximum stable amplitude that does not affect $DA_{min}$ (from [36]).*

It is rather uncommon that for a given angle the stable amplitude features considerable variations from seed to seed, so that its distribution over seeds features outliers. For our purposes, the outlier identification is carried out in steps [36]. Two types of ML approaches for the automatic detection of outliers have been tested. When using SL, the task of outlier detection is treated as a classification problem and a SVM is trained to distinguish between normal and abnormal points.

Two UL approaches have also been considered, namely the Density-Based Spatial Clustering of Applications with Noise (DBSCAN) method [60], and the Local Outlier Factor (LOF) [61] algorithm. Figure 10 reports the comparison between SVM, DBSCAN, and LOF algorithms. The labels predicted by DBSCAN and LOF were combined through a binary OR operation to create a fourth set of labels. A fifth set of labels is generated by detecting and removing false positives using a statistical method [36]. An iterative approach is performed with this post-processing, namely the algorithm starts at the minimum (maximum) point, and moves outwards (inwards), recalculating the statistical variables of the regular points at each step.

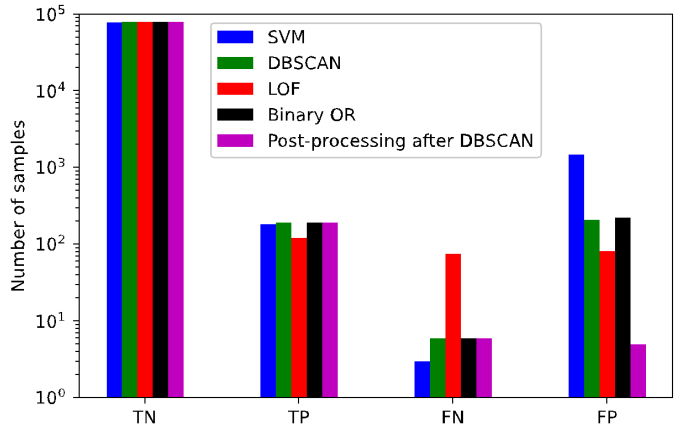

*Figure 10: Outcome of the anomaly detection performed with SVM, DBSCAN, LOF, a binary OR between DBSCAN and LOF, and post-processing following DBSCAN methods. TP = True Positives (anomaly correctly detected), TN = True Negatives (normal point correctly detected), FP = False Positives, FN = False Negatives (from [36]).*

The results indicate that the unsupervised methods have a performance an order of magnitude better than SVM in terms of false positives. However, they are worse in terms of false negatives, particularly when using LOF. The main explanation of this observation is in how the algorithm works. In supervised learning, the classes are assumed to be balanced between normal and abnormal points, as the algorithm tries to learn a decision boundary that separates the two sets of points. An imbalance induces a bias in the position of the decision boundary, so false positives are more likely. On the other hand, the unsupervised methods try to identify points further away or in lower density from other clusters of normal points, so they are expected to work better overall. The post-processing method following DBSCAN, represents an improvement as it reduces the number of false positives, whilst retaining the same TP and FN rates. The two examples in Fig. 9 are those where the classification was performed using a post-processing after DBSCAN.

As a final remark, we would like to mention that ML approaches have recently been applied to the problem of improving the quality of the model describing the evolution of DA with time [62]. In this case, Gaussian processes have been very successfully used to generate synthetic points representing the DA at a given number of turns. These synthetic points have been added to those generated by the numerical simulations, and the whole data set has been used to determine the model describing the time evolution of the DA. In this way a sizable improvement in the quality of the fit has been shown [62].

**Conclusions and outlook**

A selection of ML applications from a variety of domains linked with beam measurements for the CERN LHC have been presented and discussed in detail. In all applications, ML is seen as an efficient way to perform various types of classification processes on large data sets.

One of the first LHC ML applications originated from the quest to improve optics measurement and correction by detecting faulty BPMs. Anomaly detection together with clustering techniques were instrumental to achieving successful data cleaning. This was followed by basic Neural Network implementation for optics correction which is already producing very promising results. Further improvements are anticipated using an autoencoder network for improving the quality of betatron-phase measurements and performing noise reduction on turn-by-turn data.

For the automated alignment of collimators, excellent results have been obtained thanks to ML techniques which were used to discriminate between true beam-loss spikes and spurious events. This has resulted in a remarkable shortening of the setup time for the collimation system, with the ML implementation now fully operational. The next step is to move to the analysis of crosstalk effects between the loss signals of the two beams, as this could open the possibility to perform parallel collimator alignment with two circulating beams.

Promising results have also been obtained by using a ML model for optimising LHC beam lifetimes. However, rather than using only the operational data, this necessitated that the parameter space was properly explored by means of dedicated machine experiments. Although the ML-based model was found to describe operational configurations rather well, non-standard configurations cannot be predicted, which indicates a certain lack of predictive power. Hence, exploring alternative ML approaches is a future line of research.

Detection and classification of beam instabilities by means of Unsupervised Learning has been applied to the data collected by the LHC ObsBox system, with the goal of selecting interesting instability events from a huge collection of less interesting data. The combination of anomaly-detection techniques and Hierarchical Clustering algorithm has led to promising result. The possibility to apply such an ML approach to the actual automatic ObsBox triggering will now be explored, to reduce the amount of uninteresting data saved to file.

Possible heating issues during LHC operation can be identified by ML techniques through the classification of vacuum gauge measurements. Deep learning approaches are under investigation to provide a boost in the performance of classification algorithms following the promising initial results achieved. The ultimate goal is the deployment of this approach when operation will be resumed for the LHC Run 3.

Considering the data obtained from numerical simulation as a testbench for ML techniques has also been discussed, with the case considered that of determining outliers in the DA distribution for different configurations of magnetic field errors. An efficient approach has been devised based on Unsupervised Learning. Voting between two different clustering algorithms has produced a visible improvement in the outlier detection. However, a careful post-processing of the results obtained with DBSCAN has generated the best performance. The lowest number of false negatives has been obtained by using Supervised Learning, but the creation of ten times as many false positives, is severely hampering its usability. Finally, a successful improvement in the quality of the fit procedures used to model the DA data as a function of turn number has been achieved by means of Gaussian processes that are used to generate synthetic DA data.

## Acknowledgments

The LHC team of the Operations Group of the CERN Beams Department is warmly acknowledged for the help during the experimental sessions. The Swiss Accelerator Research and Technology institute (CHART) supports the EPFL studies.

Pasquale Arpaia

Pasquale Arpaia earned a Master Degree and PhD in Electrical Engineering at University Federico II, Napoli, Italy, where he is full professor of Instrumentation and Measurements. He is Director of the Interdepartmental Center for Research on Management and Innovation of Health (CIRMIS), Head of the Instrumentation and Measurement for Particle Accelerators Laboratory (IMPALab), and the Augmented Reality for Health Monitoring Laboratory (ARHeMlab). The main research interests include instrumentation and measurement for magnets, advanced materials, beam, superconductors, power converters, and cryogenics of particle accelerators, biomedical instrumentation, Augmented Reality, Brain Computer Interfaces, evolutionary diagnostics, distributed measurement systems, ADC modelling, and testing.

Gabriella Azzopardi

Gabriella Azzopardi graduated in Computer Science in 2016 and completed her PhD at the University of Malta in 2019. During her PhD she automated the alignment of more than 100 collimators in the Large Hadron Collider by introducing machine learning in LHC operation, speeding up collimator alignments by 70%. Currently, she is the primary developer of collimation controls at CERN, to further automate and upgrade the system in preparation for LHC startup.

Frédéric Blanc

Frédéric Blanc obtained a PhD in experimental particle physics in 1998 from the University of Fribourg, Switzerland. His research is focused on the study of heavy flavour phenomena and the indirect search for new physics through precision measurements. Expertise in multivariate analysis techniques and work on detector developments brought him to collaborate on accelerator beam losses and instabilities studies.

Xavier Buffat

Xavier Buffat studied physics at the Ecole Polytechnique Fédérale de Lausanne where he obtained a PhD in 2015. He joined CERN's Operations Group in 2009, then moved to the Accelerator and Beam Physics Group in 2011 and became a staff member of the Organisation in 2016. His research is focused on the collective forces in charged particle beams and their impact on the quality of the beams in the CERN accelerator complex.

Loic Coyle

Loic Coyle graduated in Nuclear Engineering in 2018 from Grenoble INP - Phelma and is currently doing a PhD at EPFL/CERN developing novel, machine learning based, methods to analyse and model beam losses and instabilities in the Large Hadron Collider.

Elena Fol

Elena Fol graduated in Computer Science in 2018 and earned her PhD in Physics from Goethe University in Frankfurt in 2021. During her PhD research, she worked at CERN on the development of novel methods based on machine learning for beam optics measurements and corrections. Currently, she is a senior fellow at CERN, working on the muon collider project, extending her expertise to modelling and optimisation of particle accelerators by applying machine learning algorithms.

Francesco Giordano

Francesco Giordano graduated in Electronics Engineering in 2017 and has received a PhD in Information Technology and Electrical Engineering in 2021. During his PhD research Francesco has been working at CERN in the Beam department. His areas of expertise are machine learning and beam-induced heating.

Massimo Giovannozzi

Massimo Giovannozzi has received a PhD in nonlinear beam dynamics from the University of Bologna, Italy, in 1993 and after this he joined the Accelerator Physics Group of the CERN PS division, and in 1997 he became staff. Currently, he is Principal Accelerator Physicist in the Accelerator and Beam Physics Group of the CERN Beams department and his main research activities are nonlinear beam manipulations, dynamic aperture, and diffusive models of nonlinear beam dynamics.

Tatiana Pieloni

Tatiana Pieloni has obtained her PhD in Physics from the Ecole Polytechnique Fédérale de Lausanne (EPFL), Switzerland and CERN in 2008. After a fellowship at the Paul Scherrer Institute on RF design, she accepted a staff position at CERN from 2010 till 2016 to work on the beam-beam and collective effects of the Large Hadron Collider. In 2016 she joined the Laboratory of Particle Accelerator Physics at the EPFL as a lecturer and researcher, where she teaches and conducts research in collective effects in beam dynamics.

Roberto Prevete

Roberto Prevete received a PhD in applied mathematics and computer science in 1999. He is now a researcher in Computer Science at the Department of Electrical Engineering and Information Technology (DIETI) of the University of Naples Federico II. His main research interests include machine learning and statistical methods applied to data analysis, computational neuroscience and computational modelling of biological information processing. He was responsible for the research in several national and international projects.

Stefano Redaelli

Stefano Redaelli graduated in Physics at the University of Milan, Italy, and received his PhD in accelerator physics at the University of Lausanne, Switzerland. He joined the Accelerator Physics Group at CERN in 2004, working on beam collimation aspects for the Large Hadron Collider. He then moved to the Operations Group in 2006 as an Engineer in Charge of the LHC. In 2012, he joined the Accelerator Physics group again, leading the beam collimation studies for LHC and its HL-LHC upgrade. He acted as LHC machine coordinator during the Run 2 (2015-18) participating in the performance of the accelerator at 6.5TeV.

Belen Salvachua

Belen Salvachua earned her PhD degree from Universitat de Valencia, Valencia, Spain in 2008 developing real time energy reconstruction algorithms in the ATLAS Hadronic Tile Calorimeter and exploring the ATLAS sensitivity to observation of extra-dimension signatures. She continued these studies at the Argonne National Laboratory with a Director's Fellowship from 2008 until 2011, when she was awarded with a CERN Fellowship on accelerator physics to work on the Large Hadron Collider collimation system and machine performance. In 2014, Belen became part of the CERN staff as Engineer in Charge of the operation of the CERN Large Hadron Collider (LHC). Currently she is the physicist responsible for the Beam Loss Monitoring system of the LHC.

Benoît Salvant

Benoît Salvant graduated from Ecole Supérieure de Physique et Chimie Industrielles de la Ville de Paris (ESPCI) ParisTech, France in 2003, with a Joint Honours year at Imperial College London(UK). He gained professional experience in several countries and several fields of science and technology, but in 2006 started a PhD in accelerator physics with Ecole Polytechnique Fédérale de Lausanne (EPFL), Switzerland and CERN. Benoît joined CERN as a staff member in 2012, and his task has been the follow up of beam coupling impedance in all CERN circular accelerators and became a member of the Super Proton Synchrotron supervising team. He is currently chairman of the CERN Impedance Working Group.

Michael Schenk

Michael Schenk earned his PhD in 2019 from École Polytechnique Fédérale de Lausanne (EPFL), Switzerland, for developing a novel Landau damping technique for beam instability mitigation in circular colliders. After his postdoc at EPFL, where he worked on machine learning for beam lifetime prediction, he is now a senior fellow at CERN, where his main activities lie in developing accelerator controls applications based on reinforcement learning within the Operations Group.

Matteo Solfaroli Camillocci

Matteo Solfaroli graduated in Physics at the University of Rome, Italy. He joined the LHC hardware commissioning team at CERN in 2007, working on the technical coordination. He then moved to the Operations Group in 2009 as CERN staff. As an Engineer in Charge, he operated the LHC in Run 1 and Run 2, contributing to the performance optimisation. During the same period he was also in charge of the recommissioning of the LHC superconducting circuits. He then became a machine Physicist in 2020 and will act as LHC machine coordinator during the Run 3.

Rogelio Tomás

Rogelio Tomás received a PhD in 2003 from the Universidad de Valencia, Spain, on the beam-based measurement of resonance driving terms. He then moved to the Brookhaven National laboratory for a postdoc on the optimization and operation of the Relativistic Heavy ion collider. Rogelio was then appointed Staff in the Spanish light source until 2005, and since then he is staff at CERN in the Accelerator and Beam Physics Group working on optics measurement and design of current and future accelerators.

Gianluca Valentino

Gianluca Valentino received a PhD in computer engineering from the University of Malta in 2013 and subsequently was a Marie Curie COFUND postdoctoral fellow in the Accelerator and Beam Physics Group at CERN. In 2016 he joined the Department of Communications and Computer Engineering at the University of Malta as a lecturer, where he teaches and conducts research in machine learning, pattern recognition, and signal processing.

Frederik Florentinus Van der Veken

Frederik Van der Veken earned his PhD in 2014 in theoretical particle physics at the University of Antwerp, on the mathematical properties of Wilson Lines in quantum chromodynamics. He then moved into the field of accelerator physics, as a Marie Curie COFUND fellow in the Accelerator and Beam Physics Group at CERN. Afterwards he continued there as a Project Associate with the University of Malta, and from 2021 he is a staff member in the same group. His current research is mainly focused on nonlinear beam dynamics and collimation commissioning and optimisation.

Jörg Wenninger

Jorg Wenninger earned his PhD in particle physics with the L3 experiments at LEP at the University of Geneva in 1992. He then switched sides and moved from being a beam user to a beam producer in the Beam Instrumentation and Operations Groups at the SPS, LEP, and LHC accelerators. He is currently leading the LHC beam operation team at CERN.